Tuning edge localized spin waves in magnetic microstripes by proximate magnetic structures

Zhizhi Zhang[1,2], Michael Vogel[1], M. Benjamin Jungfleisch[3], Axel Hoffmann[1], Yan Nie[2,*], and Valentine Novosad[1,*]

[1]Materials Science Division, Argonne National Laboratory, Argonne, IL 60439, USA
[2]School of Optical and Electronic Information, Huazhong University of Science and Technology, Wuhan 430074, China
[3]Department of Physics and Astronomy, University of Delaware, Newark, DE 19716, USA

**Abstract**

The propagation of edge localized spin waves (E-SWs) in yttrium iron garnet (YIG) microstripes with/without the proximate magnetic microstructures is investigated by micromagnetic simulations. A splitting of the dispersion curve with the presence of permalloy (Py) stripe is also observed. The E-SWs on the two edges of YIG stripe have different wavelengths, group velocities, and decay lengths at the same frequencies. The role of the Py stripe was found to be the source of the inhomogeneous static dipolar field without dynamic coupling with YIG. This work opens new perspectives for the design of innovative SW interference-based logic devices.

**Introduction**

Data transmission with spin waves (SWs) and their particle-like analog magnons is a promising direction for the next generation information devices because of their low heat dissipation and high efficiency[1-5]. One important application of SWs is logic operations that are based on the interaction of waves, especially wave interference[6-10]. Therefore, SWs need to be wave vector monochromatic and well localized. Under this prerequisite, a certain number of researches focus on SWs in domain walls[11-13]. However, the frequencies of the classic domain wall SWs are lower than those in domains. Besides, the SWs propagating in microstripes can reach several GHz frequencies. There are two kinds of SWs in microstripes: waveguide SWs (W-SWs) and edge localized SWs (E-SWs). The W-SWs contain a set of multiple modes with various wave vectors hybridized in the central region of the microstripes[14-16]. In contrast, the E-SWs are well confined in narrow channels at the edges of the microstripes[17-20]. Their wave vector components are also monochromatic. Thus, an effective method for manipulating the propagation of E-SWs is a critical step toward the development of E-SW based magnonic devices[21].

In this work, we studied the propagation of the E-SWs in yttrium iron garnet (YIG) stripes without/with laterally proximate permalloy (Py) stripes close to one edge. YIG has the lowest known damping factor ($\alpha$) and magnetization saturation ($M_s$) compared with the metal material.[22] The material Py was selected for its $M_s$, which is almost six times larger than $M_s$ of YIG, while $\alpha$ maintains a low value. We calculated dispersion diagrams of the SWs in the waveguides and quantitatively analyzed the E-SWs in YIG stripes, including the wavelengths ($\lambda$), decay lengths ($\delta$), and group velocities ($v_g$) at certain frequencies. Furthermore, the effects of the Py stripe on the E-SWs in YIG stripe were fully explored, and the mechanisms of our findings were revealed, potentially having implications for future engineering applications.

**Methods**

We performed micromagnetic simulations on SWs propagating in magnetic thin-film microstrips using MuMax3[23]. Fig. 1 (a) shows the schematic of the studied model: a 10 μm × 1 μm × 50 nm YIG microstripe and a proximate Py stripe with the same sizes laterally close to one edge (YIG/Py). A global external magnetic field ($H_{ext}$) of 1000 Oe in the y-direction was applied to the structure, corresponding to the Damon-Eschbach (DE) geometry[24]. Material parameters used in the simulation were $M_s$ = 1.48×10$^5$ A/m, exchange constant $A_{ex}$= 4×10$^{-12}$ J/m, and $\alpha$ = 7.561×10$^{-4}$ for YIG[25] and $M_s$ = 8.6×10$^5$ A/m, $A_{ex}$= 1.3×10$^{-11}$ J/m, $\alpha$=0.01 for Py. In addition, the attenuating areas (4 μm on each end, not shown here) with $\alpha$ gradually increased to 0.25, served as the absorption boundaries to avoid the reflection at the ends of the stripes to simulate the case of infinitely long stripes.

To analyze the dispersion relations of the propagating SWs in the magnetic stripes, the excitations were applied locally in the antenna area using a sine cardinal function $h_x = h_0 \dfrac{\sin(2\pi f_c (t-t_0))}{2\pi f_c (t-t_0)}$ [26] with a cut off frequency $f_c$ = 50 GHz (Fig. 1 (b)) and $h_0$ = 10 Oe. The $f_c$ was high enough to satisfy the SW propagating conditions of both YIG and Py, and the $h_0$ was low enough to avoid nonlinear effects.[27] Fig. 1 (b) shows the sine cardinal excitation in time domain, which has a pulse-like shape. The frequency spectrum obtained by fast Fourier transform (FFT) shown in Fig. 1 (c) indicates that such kind of excitation has a uniform intensity in the whole frequency band under $f_c$ and zero intensity above $f_c$. The total simulation time was 200 ns, and the result recorded the dynamic normalized magnetization ($m_z/M_s$) evolution with the time and the length of YIG stripe. Therefore, the $m_z/M_s$ is a two-dimensional matrix. The dispersion relations were obtained through the two-dimensional FFT (2D-FFT) operation on the $m_z/M_s$.[28] Subsequently, continuous sine excitations with specific frequencies were applied continuously to study the detailed SW properties ($\lambda$, $v_g$ and $\delta$). The $H_{eff}$ extracted from the simulation was used for further analysis, as shown below.

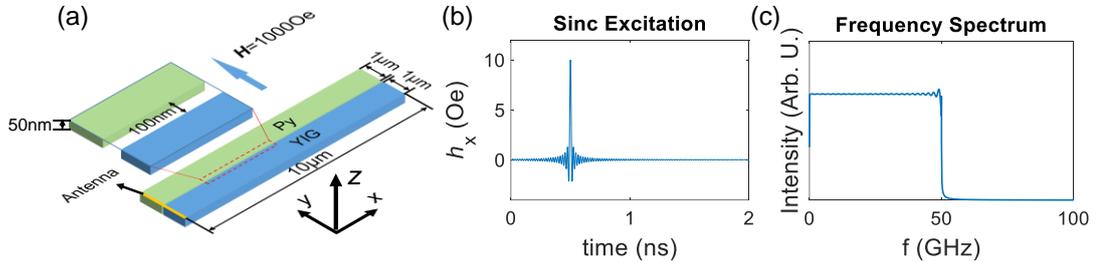

Fig. 1 (a) Schematic of proposed YIG/Py structures. The yellow antenna region indicates the SW generation. (b) Temporal evolution of sine cardinal function field with $h_0$ = 10 Oe applied along x-axis with a Gaussian distribution centered in yellow antenna region. (c) Frequency spectrum obtained from FFTs of applied sine cardinal field with $f_c$ = 50 GHz.

**Discussion**

The dispersion diagrams of the single YIG and the YIG and Py stripes in YIG/Py are shown in Fig. 2 (a), (b), and (c), respectively. In all stripes, there clearly exist a set of hybridized W-SWs localized in higher frequency band and E-SWs in lower frequency band. According to Ref. [29], the SWs dispersion relation of DE geometry in lossless materials can be theoretically written as

$$f = \gamma \sqrt{H_{eff}(H_{eff} + M_s) + \frac{M_s^2}{4}(1 - e^{-2kd})} \quad (1)$$

where $d$ is the thickness of the film, $k$ is the wave vector, $\gamma$ is the gyromagnetic ratio (2.8 MHz/Oe), $\lambda_{ex}$ is the exchange length and is equal to $\sqrt{A_{ex}/2\pi M_s^2}$ (in CGS)[30]. The formation of E-SWs is the prominently reduced $H_{eff}$ at the edges by the demagnetization field, leading to the lower frequencies than those of W-SWs[31]. In addition, the E-SWs on the two edges of a single YIG stripe have a degenerated dispersion curve because of the symmetric magnetic configurations. In contrast, the degenerated states were split into two curves, as shown in Fig. 2 (b), which is accompanied by the presence of Py. Because of the much higher $M_s$ of Py compared to YIG, the SWs in Py stripe propagate at remarkably higher frequencies than those of YIG under the same $H_{ext}$, as shown in Fig. 2 (c) and in accordance with Eq. (1).

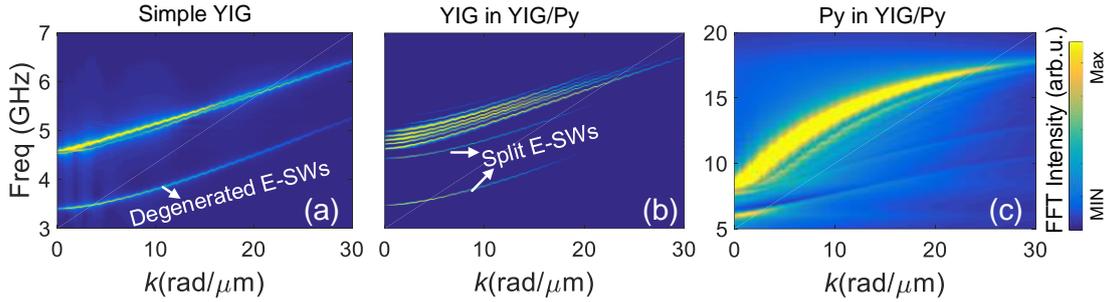

Fig. 2 Dispersion relations for both W-SWs and E-SWs propagating in (a) single YIG stripe without Py stripe, (b) YIG stripe in YIG/Py, and (c) Py stripe in YIG/Py. $k$ is along the $x$-axis.

The dispersion diagrams of YIG stripes without/with proximate Py stripe were zoomed in for further analysis, as shown in Fig. 3 (a) and (b), respectively. To study the detailed behaviors of the E-SWs propagating on the two edges of YIG stripes, we applied a continuous excitation of the sine function $h_x = h_0 \sin(2\pi f t)$ in the antenna region with $f = 3.9$ and 4.5 GHz, respectively. Here, $h_0 = 1$ Oe is weak enough to avoid nonlinear effects.[27] The total simulation time was 80 ns to ensure that the system reaches a steady state. To obtain the accurate values of $\lambda$ and $\delta$, we fit the $m_z/M_s$ space distributions in steady state ($t$ = 80 ns) using the following equation:

$$\frac{m_z}{M_s} = A \sin\left(\frac{2\pi}{\lambda} x + \theta\right) \exp\left(-\frac{x}{\delta}\right) \quad (2)$$

where $\theta$ is a phase factor and $A$ is a scaling factor. The $v_g$ of every E-SW was determined using $l/\tau$, where $l$ is the length of the stripe and $\tau$ is the time for the $m_z/M_s$ at the right end reaching to the stable state (see Fig. 3 (f)). Fig. 3 (c) and (d) show the $m_z/M_s$ of the single YIG and the YIG in YIG/Py on both edges at $t$=80 ns under the 3.9-GHz excitation. These figures indicate that the 3.9-GHz E-SWs propagate on both edges of the single YIG stripe with the same $\lambda$, $\delta$, and $v_g$ (Supplementary Movie 1[32]). In contrast, these E-SWs can propagate only on the edge farther away from Py in the YIG stripe. On the edge closer to Py, the oscillation of the $m_z/M_s$ was confined near the excitation (Supplementary Movie 2[32]). The 4.5-GHz E-SWs can propagate on both edges of the YIG stipe in YIG/Py, as shown in Fig. 3 (e), but with different $\lambda$, $\delta$, and $v_g$ (Supplementary Movie 3[32]). The time evolution of $m_z/M_s$ shown in Fig. 3 (f) indicates the spin waves propagation process includes the transient state (yellow region) and the steady state (green region).

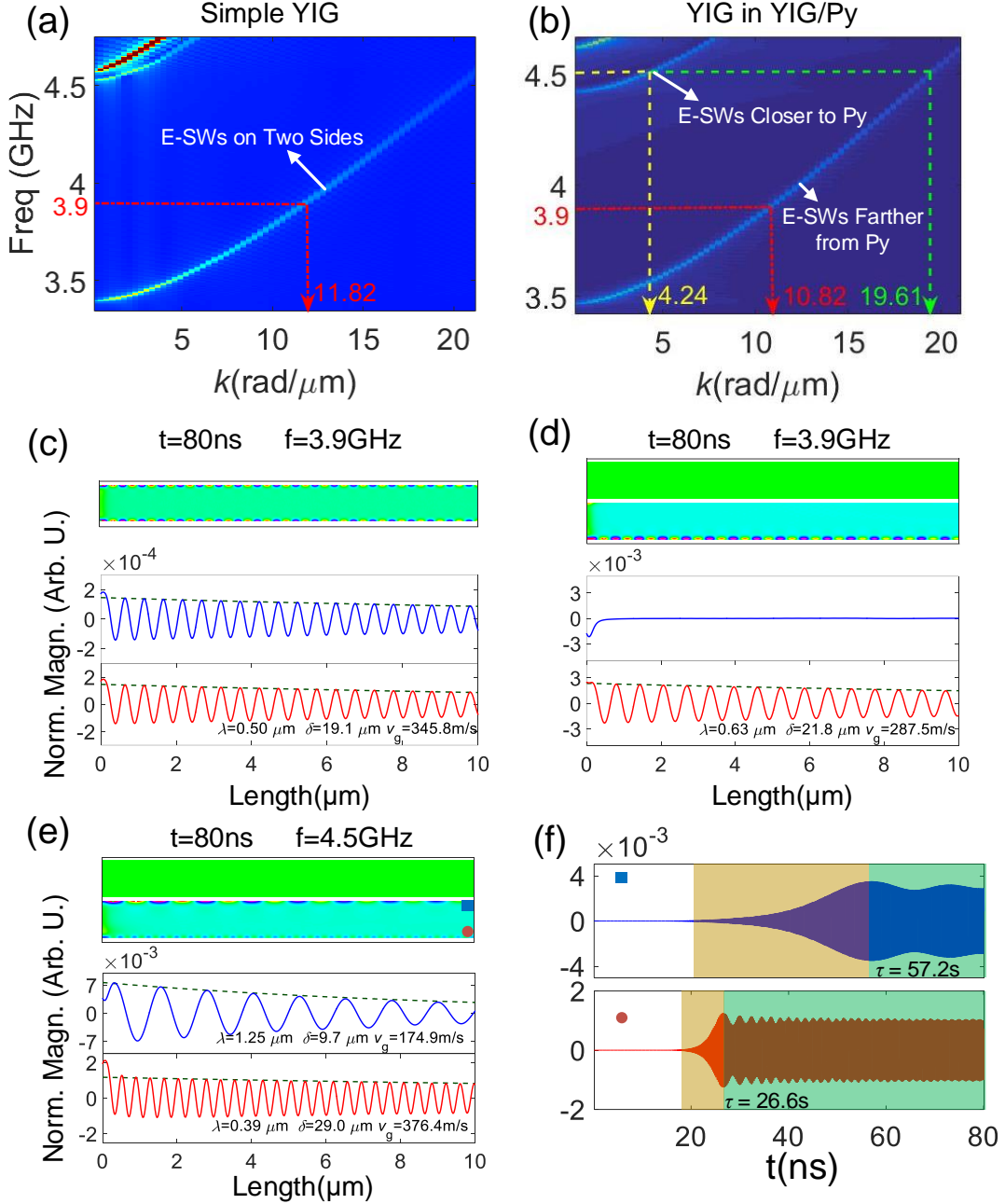

Fig. 3 Zoomed in dispersion diagram focused on E-SWs in (a) single YIG and (b) YIG stripe in YIG/Py. The dashed lines depict the wave vectors for specific excitation frequencies. Response of the $m_z/M_s$ in (c) single YIG, (d) YIG stripe in YIG/Py at t = 80 ns under 3.9-GHz excitation and (e) YIG stripe in YIG/Py at t = 80 ns under 4.5-GHz excitation. Upper panels

are the global 2D maps of the $m_z/M_s$ intensity (Green represents Py, same hereinafter); lower panels are the $m_z/M_s$ intensity distribution along the two edges of YIG stripes, where the blue curves are for the edge closer to Py, the red curves are for the edge far away from Py, and the green dash curves are the envelop line obtained from the fitting. (f) The temporal evolution of $m_z/M_s$ monitored at the ends of the two edges of the YIG stripe in the case of (e). Blue square and red circle dots represent the positions shown in upper panel of (e). The yellow region in the plot indicates the transient states, and the green region indicates the stable states. $\tau$ is the time for the $m_z/M_s$ at the positions reaching to the stable state.

To comprehensively understand the striking difference of the E-SWs propagating in the two edges of YIG stripe in YIG/Py, the impacts of the proximate Py on the YIG stripe were inspected from two aspects: the static $H_{eff}$ varied by Py and the dynamic coupling effect between YIG and Py.

The $H_{eff}$ across a single YIG stripe under the 1000 Oe applied field is shown in Fig. 4 (a). The figure shows that the $H_{eff}$ distribution has the following features: 1. Because of the demagnetizing field distributed at the edges of the stripe, the $H_{eff}$ is weaker than the applied field; 2. Inside the YIG stripe, the $H_{eff}$ reduces significantly close to the edges, creating the SW potential wells[33], where the dynamic magnetization is confined similar to the localization of quantum particles in potential wells; 3. The SW potential wells in the two edges of the single YIG stripe have a symmetric profile because of the symmetric geometry of the magnetic system. In contrast, the $H_{eff}$ distributions in YIG and Py stripes in YIG/Py shown in Fig. 4 (b) indicate it is significantly changed by the magnetic dipoles in Py. The depths and positions of the SW potential wells on the two edges of single YIG (from 320 to 970 Oe) are similar with those (from 340 to 930 Oe) on the edge farther away from Py of the YIG in YIG/Py. The increase of $H_{eff}$ results in the increase of $\lambda$[34, 35] and the decrease of $v_g$ and $\delta$[36], agreeing with the results in Fig. 3.

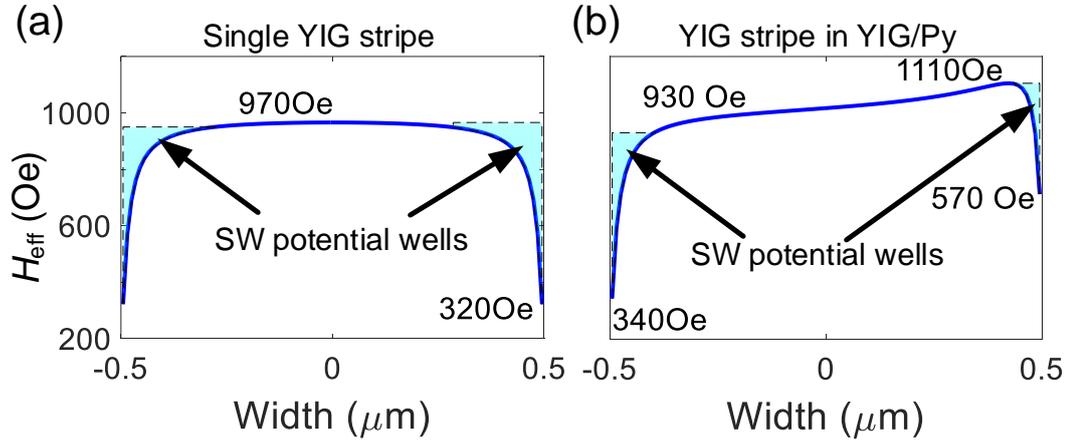

Fig. 4 The $y$-component of $H_{eff}$ across (a) single YIG stripe and (b) YIG stripe and Py stripe in YIG/Py under 1000 Oe.

To quantitatively analyze the additional field introduced by Py, we numerically calculated the dipolar field induced by Py stripe ($H_{dip-Py}$), as shown in Fig. 5 (a), and fitted it using the following equation:

$$H_{fit} = \frac{a}{r^n} + b \qquad (3)$$

where $a$ is a real scaling parameter, $b$ is an offset, $r$ is the distance to the edge of the Py stripe, and $n$ is the coefficient to be determined. $H_{\text{dip-Py}}$ was found to be proportional to $1/r$, whose intensity was about 280 Oe at $r = 0.1$ μm and reduced rapidly to 20 Oe at $r = 1.1$ μm. For further comparison, the dispersion relations of the YIG stripe in YIG/Py with $d_{\text{gap}} = 200$ and 1000 nm were numerically calculated respectively, as shown in Fig. 5 (b) and (c). We noticed that the curves of the E-SWs dispersions were getting closer with the increasing of the gap distance compared with Fig. 2(b), and finally, they almost merged together at $d_{\text{gap}} = 1000$ nm, where the $H_{\text{dip-Py}}$ decayed to nearly zero. In a brief summary, the results show that the static $H_{\text{dip-Py}}$ introduced by Py was one factor resulting in the difference of the E-SWs propagating on the two edges of the YIG stripe in YIG/Py.

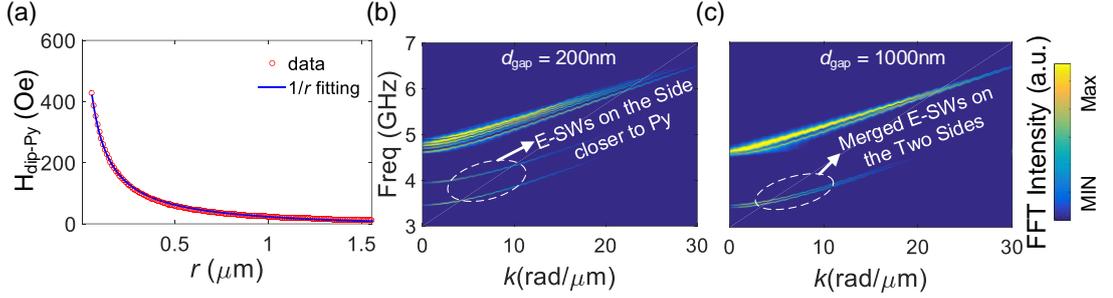

Fig. 5 (a) Profile of the $y$-component $H_{\text{dip-Py}}$ versus the distance to the fully magnetized Py stripe. (b) The dispersion relations of the SWs propagating in YIG stipes in YIG/Py with $d_{\text{gap}}$ = 200 and 1000 nm, respectively.

If the two magnetic structures were close to each other, dynamic coupling might occur. Dynamic coupling means the dynamic magnetization transfers repeatedly between one structure and the other, resulting in the splitting of the dispersion curves[5]. The dynamic coupling has been observed in two YIG stripes horizontally close to each other[5, 37-39] and multiple layers vertically with different materials[40-43]. In those cases, the frequencies of the SWs in different magnetic structures overlapped with each other under certain conditions. In our case, according to the dispersion relations shown in Fig. 2(b) and (c), the frequencies of the propagating E-SWs in YIG and Py were not overlapped. Therefore, no dynamic coupling occurred during the propagation of the studied E-SWs. To further confirm this point, we perform a simulation on the single YIG stripe under 4.5-GHz excitation. The static external field was set as the superposition of the 1000 Oe homogeneous field and the inhomogeneous field described by Eq. (3) with $n = 1$ in the width direction. The 2D maps of the $m_z/M_s$ intensity in YIG stripe and the temporal evolution of $m_z/M_s$ on the two edges at the end of the YIG stripe are shown in Fig. 6 (a) and (b). The E-SWs in this case showed the same behaviors as those shown in Fig. 3 (e) and (f). Consequently, the role of Py stripe is simply a source of an additional inhomogeneous magnetic field without dynamic coupling with YIG.

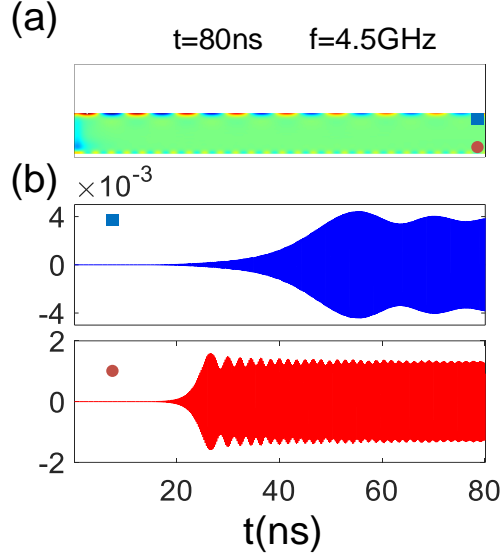

Fig. 6 (a) 2D maps of the $m_z/M_s$ intensity of the YIG stripe ($f$ = 4.5 GHz and $t$ = 80 ns) under the static external field consisting of the 1000 Oe homogeneous field and the inhomogeneous field aligned perpendicular to the long side of the stripe, as described by Eq. (3) with n = 1. (b) The temporal evolution of $m_z/M_s$ monitored at the end of the YIG stripe. Blue square and red circle dots represent the corresponding positions shown in (a).

Here, by tuning the position of the dispersion curve in the diagram, the SWs can propagate with designed properties in two separated channels in just one waveguide. In addition, the induced static dipolar field is an essential factor for tuning the dispersion curves. It strongly depends on the magnetization of the proximate magnet. Consequently, we can expect to actively tune the E-SWs by changing the magnetization, for example, the temperature of the proximate Py stripe can be controlled by applying the charge current. The Joule heat changes the magnetization as well as the induced dipolar field. Such performances indicate a method toward producing novel magnonic devices.

**Conclusion**

In summary, we studied the E-SWs propagating behaviors on the two edges of the YIG stripe with/without the laterally proximate Py stripe. The degenerated dispersion curve of the E-SWs in YIG stripe was split into two curves with the presence of the Py stripe. Correspondingly, the E-SWs on the two edges of YIG stripe have different $\lambda$, $v_g$, and $\delta$ at the same frequencies. The reasons for the splitting of the dispersion curve were investigated through exploring the role of the Py played in the magnetic structure. The additional Py stripe acts as a source of the inhomogeneous magnetic field. No dynamic coupling occurred between YIG and Py in the structure. The results show that unique characteristics of SWs integrated in a single waveguide opens perspectives for the design of innovative logic elements based on constructive or destructive SW interference.

**Acknowledgment**
Work at Argonne was supported by the U. S. Department of Energy, Office of Science, and

Materials Science Division. Use of the Center for Nanoscale Materials was supported by the U. S. Department of Energy, Office of Science, Basic Energy Science, under contract no. DE-AC02-06CH11357. Zhizhi Zhang acknowledges the financial support of China Scholarship Council (no. 201706160146). The authors thank Prof. Andrii V. Chumak, Prof. Paul A. Crowell and Dr. Yi Li for constructive discussions.